\documentclass[12pt]{article}
\usepackage{epsfig}

\def\beq{\begin{equation}}
\def\eeq#1{\label{#1}\end{equation}}
\def\eeqn{\end{equation}}

\def\beqa{\begin{eqnarray}}
\def\eeqa#1{\label{#1}\end{eqnarray}}
\def\eeqan{\end{eqnarray}}

\let\bar=\overbar

\def\ie{{\it i.e.}}

\def\Dslash{\not{\hbox{\kern-4pt $D$}}}
\def\dslash{\not{\hbox{\kern-2pt $\del$}}}

\def\msb{{\bar{\ssstyle M \kern -1pt S}}}

\def\Title#1{\begin{center} {\Large {\bf #1} } \end{center}}

\begin{document}

\Title{Embedding Dark Energy in Supergravity}

\bigskip\bigskip


\begin{raggedright}

{\it Philippe Brax\\
Service de Physique Th\'eorique, CEA/DSM/SPhT, Unit\'e de
recherche associ\'ee au CNRS, CEA--Saclay 91191 Gif/Yvette cedex,
France\\}
\bigskip\bigskip
\end{raggedright}

Dark energy\cite{IA} is one of the most intriguing puzzles of
present day physics. When interpreted within the realm of General
Relativity, its existence is linked to the presence of a weakly
interacting fluid with a negative equation of state and a dominant
energy density. The simplest possibility is of course a pure
cosmological constant. A plausible alternative involves the
presence of a scalar field responsible for the tiny vacuum energy
scale\cite{RP,quint,ed}.  In most cases, the quintessence field
has a runaway potentials and takes large values now, of the order
of the Planck mass. This suggests to  embed such models in high
energy physics\cite{PB,BMcosmo}. The most natural possibility is
supergravity as it involves both supersymmetry and gravitational
effects. Moreover, superstring theories lead to supergravity
models at low energy.

\par

From the model building point of view, the quintessence field does
not belong to  the standard model. Hence there must be a separate
dark energy sector. The observable sector is well-known and the
hidden supersymmetry (SUSY) breaking sector can be
parameterised\cite{Nilles}. In the following, we give a brief
overview of some of the constraints on the embedding of  dark
energy in broken supergravity following mostly~\cite{BM1,BM2,BM3}.

\section{Coupling Dark Energy to SUSY Breaking}

\par

As soon as a quintessence field has a runaway potential and leads
to the present day acceleration of the universe expansion, its
mass is tiny and may lead to gravitational problems. In order to
minimise this problem, we assume that the quintessence sector is
only coupled gravitationally to the observable and hidden sectors.
This can be described by the K\"ahler and super potentials
\begin{equation}
K= K_{\rm quint} + K_{\rm hid} + K_{\rm obs},\ W= W_{\rm quint} +
W_{\rm hid}+  W_{\rm obs}\, .
\end{equation}
The observable sector comprises the fields of the Minimal Standard
Supersymetric Model (MSSM) $\phi^a$ and the corresponding
superpotential can be expressed as
\begin{equation}
W_{\rm obs}= \frac{1}{2}  \mu_{ab} \phi^a \phi^b +\frac{1}{3}
\lambda_{abc} \phi^a \phi^b \phi^c \, ,
\end{equation}
where $\mu_{ab}$ is a supersymmetric mass matrix and
$\lambda_{abc}$ the Yukawa couplings.

\par

SUSY breaking causes the appearance of soft terms in the
observable and dark sectors. We can parameterise the hidden sector
supersymmetry breaking in a model independent way
\begin{equation}
\label{parahidden} \kappa ^{1/2}<{z_i}>_{\rm min}\sim a_i(Q)\, ,
\quad \kappa
    <{W_{\rm hid}}>_{\rm min}\sim M_{_{\rm S}}(Q)\, , \quad \kappa
    ^{1/2}<{\frac{\partial W_{\rm hid}}{\partial z_i}}>_{\rm
    min}\sim c_i(Q)M_{_{\rm S}}(Q)\, ,
\end{equation}
where $a_i$ and $c_i$ are coefficients of order one which depend
on the detailed structure of the hidden sector, $M_{_{\rm S}}$ is
the SUSY breaking scale  and $\kappa \equiv 8\pi /m_{\rm pl}^2$.
Notice that the coupling of the hidden sector to quintessence
implies that the vev's of the hidden sector fields $z_i$
responsible for supersymmetry breaking can depend on the
quintessence field. The observable potential reads
\begin{eqnarray}
V_{_{\rm mSUGRA}} &=&\cdots + {\rm e}^{\kappa K}V_{\rm
susy}+e^{\kappa K}{\cal A}(Q)\lambda_{abc}\left(\phi_a \phi _b\phi
_c+\phi_a^{\dagger} \phi _a^{\dagger }\phi _c^{\dagger }\right)
+e^{\kappa K}B(Q) \mu_{ab}\left(\phi _a \phi _b+\phi _a ^{\dagger
}\phi _b^{\dagger }\right) \nonumber \\ &+& m_{a\bar{b}}^2\phi _a
\phi _{b}^{\dagger }\, .
\end{eqnarray}
where the soft terms are the terms which are not in $V_{\rm
susy}$.

 We consider that the dark energy superpotential is of the
form
\begin{equation}
W_{\rm quint}(Q)\equiv M^3{\cal W}\left(\kappa ^{1/2}Q\right)\, .
\end{equation}
where $M$ is a scale characterising dark energy. The choice of the
K\"ahler potential is also crucial. As an example, we will focus
on the no-scale case corresponding to K\"ahler moduli
\begin{equation}
K_{\rm quint} =-\frac{3}{\kappa }\ln \left[\kappa
  ^{1/2}\left(Q+Q^{\dagger }\right)\right]\, ,
\end{equation}
The kinetic terms of the moduli read $3\vert
\partial Q\vert ^2/\left(Q+Q^{\dagger}\right)^2$ implying that $Q$
is not a normalized field. The normalized field $q$ is given by
\begin{equation}
\kappa^{1/2} Q=\exp\left(-\sqrt{\frac{2}{3}} q\right)\, .
\end{equation}
where $q$ is a dimensionless scalar field.

In the no scale case and if  $W_{\rm hid}$ is constant, $M_{_{\rm
S}}$ is constant, $\cal A$ and $B$ are constant of the order of
$M_{_{\rm S}}$, and
\begin{equation}
\label{uni} 2B=-M_{_{\rm S}}+3{\cal A}\, ,
\end{equation}
while the mass $m_{a\bar{b}}$ acquires a very simple
$Q$-dependence given by
\begin{equation}
m_{a\bar{b}} =\frac{M_{_{\rm S}}}{\left[\kappa
^{1/2}\left(Q+Q^{\dagger }\right)\right]^{3/2}}\delta_{a\bar b}\,
.
\end{equation}
In general, the soft terms have a non-trivial dependence on $Q$.

We now consider the application of the previous results to the
electroweak symmetry breaking. The Higgs potential also becomes a
$Q$-dependent quantity. The total Higgs potential, taking $H_{\rm
u}^0$ and $H_{\rm d}^0$ to be  real reads
\begin{eqnarray}
\label{simpleVhiggs} V^{_{\rm Higgs}}&=& {\rm e}^{\kappa K_{\rm
quint}}\biggl[\left(\left \vert \mu \right \vert ^2 +m_{H_{\rm
u}}^2\right)\left\vert H_{\rm u}^0\right\vert ^2+ \left(\left
\vert \mu \right \vert ^2 +m_{H_{\rm d}}^2\right)\left\vert H_{\rm
d}^0\right\vert ^2-2\mu B(Q)\left\vert H_{\rm u}^0\right\vert
\left\vert H_{\rm d}^0\right\vert\biggr]\nonumber \\ & & +\frac18
\left(g^2+g'^2\right) \left( \left\vert H_{\rm u}^0\right\vert
^2-\left\vert \ H_{\rm d}^0\right \vert^2\right )^2 \, .
\end{eqnarray}
In presence of dark energy, the minimum becomes $Q$--dependent and
the particles of the standard model acquire a $Q$-dependent mass.
Straightforward calculations give
\begin{eqnarray}
\label{min1} {\rm e}^{\kappa K_{\rm quint}}\left(\left \vert \mu
\right \vert ^2+m_{H_{\rm u}}^2\right) &=& \mu B(Q)\frac{{\rm
e}^{\kappa K_{\rm quint}}}{\tan \beta
}+\frac{m_{Z^0}^2}{2}\cos \left(2\beta \right)\, , \\
\label{min2} {\rm e}^{\kappa K_{\rm quint}}\left(\left \vert \mu
\right \vert ^2+m_{H_{\rm d}}^2\right)&=&\mu B(Q){\rm e}^{\kappa
K_{\rm quint}}\tan \beta -\frac{m_{Z^0}^2}{2}\cos \left(2\beta
\right)\, ,
\end{eqnarray}
where we have defined the Higgs vevs as $\langle H_{\rm
u}^0\rangle \equiv v_{\rm u}$, $\langle H_{\rm d}^0\rangle \equiv
v_{\rm d}$, $\tan \beta \equiv v_{\rm u}/v_{\rm d}$,  and
$m_{Z^0}$ as the gauge boson $Z^0$.
\par

From the equations~(\ref{min1}) and (\ref{min2}), one can also
deduce how the scale $v\equiv \sqrt{v_{\rm u}^2+v_{\rm d}^2}$
depends on the quintessence field. This leads to
\begin{eqnarray}
\label{generalv} v(Q)=\frac{2{\rm e}^{\kappa K_{\rm
quint}/2}}{\sqrt{g^2+g'{}^2}} \sqrt{\left\vert\left\vert \mu
\right \vert ^2+m_{H_{\rm u}}^2\right\vert}+{\cal
O}\left(\frac{1}{\tan
  \beta }\right)\, .
\end{eqnarray}
in the large $\tan\beta$ regime.

\par

Then, finally, one has for the vev's of the two Higgs fields
\begin{eqnarray}
\label{vuvd} v_{\rm u}(Q)&=&\frac{v(Q)\tan \beta
(Q)}{\sqrt{1+\tan^2 \beta (Q)}}
=v(Q)+{\cal O}\left(\frac{1}{\tan ^2\beta }\right)\, ,\\
\label{vuvd2} v_{\rm d}(Q) &=& \frac{v(Q)}{\sqrt{1+\tan ^2\beta
(Q)}} =\frac{v(Q)}{\tan \beta }+{\cal O}\left(\frac{1}{\tan
^2\beta }\right)\, ,
\end{eqnarray}
at leading order in $1/\tan ^2\beta $. This allows us to deduce
the two kinds of fermion masses, depending on whether the fermions
couple to $H_{\rm u}$ or $H_{\rm d}$
\begin{equation}
m_{{\rm u},a}^{_{\rm F}}(Q)= \lambda_{{\rm u},a}^{_{\rm F}} {\rm
  e}^{\kappa K_{\rm quint}/2}v_{\rm u}(Q)\,
  , \quad m_{{\rm d},a}^{_{\rm F}}(Q)=\lambda_{{\rm d},a}^{_{\rm F}}
  {\rm e}^{\kappa K_{\rm quint}/2}v_{\rm
  d}(Q)\, ,
\end{equation}
where $\lambda_{{\rm u},a}^{_{\rm F}}$ and $\lambda_{{\rm
d},a}^{_{\rm F}}$ are the Yukawa coupling of the particle $\phi_a$
coupling either to $H_{\rm u}$ or $H_{\rm d}$. The masses pick up
a $\exp\left(\kappa K_{\rm quint}/2\right)$ dependence from the
expression of $v(Q)$ and another factor $\exp\left(\kappa K_{\rm
quint}/2\right)$ from the definition of the mass itself. As a
result we have $m\propto \exp\left(\kappa K_{\rm quint}\right) $
In no scale quintessence the behaviour of the standard model
particle masses is universal and given by $ \label{scalingmass}
m(Q)\propto \frac{1}{\left[\kappa ^{1/2}\left(Q+Q^{\dagger
}\right)\right]^3}\propto {\rm e}^{-\sqrt{6}q}\, . $

After electro-weak symmetry breaking, the low energy action in the
Einstein frame reads
\begin{equation}
S= \int d^4 x\sqrt{-g} (\frac{R}{2\kappa} -\frac{1}{2} g^{\mu\nu}
(\partial_\mu q)(\partial_\nu q)- V_{\rm DE}(q)) + S_m(\psi_u,
A^2_u(q) g_{\mu\nu}) + S_m (\psi_d, A^2_d(q) g_{\mu\nu})
\end{equation}
where $ K_{Q\bar Q} (\partial Q)^2= \frac{1}{2} (\partial q)^2$
and $ A_{u,d}(q)= \frac{m_{u,d}(q)}{m_{u,d}}$ is the ratio of the
$q$ dependent masses to their values in the absence of coupling to
dark energy. Notice that in general, the particles $\psi_{u,d}$
coupling to $H_{u,d}$ do not couple to gravity in an universal
way, hence a violation of the weak equivalence principle. In the
following, we will neglect the $q$ dependence of $m_{\rm H_{u,d}}$
and $B$ leading to $A_{u,d}(q)=A(q)$. This is exact in the no
scale case.

 If the dark energy potential is of the runaway type then this
implies that the quintessence field has a mass $m_q\sim H_0$, \ie
of the order of the Hubble rate now. The range
of the force mediated by the quintessence field is large. In order to satisfy the
constraints coming from fifth force experiments such as the recent
Cassini spacecraft experiment, one must require that the Eddington
(post-Newtonian) parameter $\vert \gamma -1\vert \le 5\times
10^{-5}$. If one defines the parameter $\alpha _{\rm u,d}$ by
\begin{equation}
\label{alpha} \alpha_{\rm u,d}(Q) \equiv \left\vert  \frac{{\rm
d}\ln m_{\rm u,d}^{_{\rm F}}(q)}{{\rm d} q} \right \vert \, ,
\end{equation}
where the derivative is taken with respect to the normalized field
$q$, then one must impose that
$\alpha_{\rm u,d}^2\le 10^{-5}$ since one has $\gamma =1+2\alpha
_{\rm u,d}^2$\cite{GR,DP}. This leads to a bound on $\alpha_{\rm
u,d}(q)\approx \frac{1}{2} \partial_q K_{\rm quint}$
\begin{equation}
\partial_q K_{\rm quint} \le 10^{-2}
\end{equation}
Notice the analogy with the $\eta$ problem of inflation. This constraint can
be satisfied  using an appropriate shift symmetry. In the no scale
case, Eq.~(\ref{scalingmass}) implies $ \alpha_{\rm u,d}=
\sqrt{6}\,  $ in contradiction with the bounds on the existence of
a fifth force.

\par

However, the above description is too naive because we have not
taken into account the chameleon effect. Indeed, in the presence
of surrounding matter like the atmosphere or the inter-planetary
vacuum, the effective potential for the quintessence field is
modified by matter and becomes\cite{KW,cham}
\begin{equation}
\label{poteff} V_{\rm eff}(Q)= V_{_{\rm DE}}(Q) + A(Q) \rho_{\rm
mat}\, ,
\end{equation}
where $A(Q)$ is the coupling of the quintessence field to matter.
This can lead to an effective minimum for the potential even
though the Dark Energy potential is runaway. The theory is
compatible with gravity tests if\cite{KW}
\begin{equation}
\frac{\alpha _q q_{\rm now}}{\Phi _{_{\rm N}}}\ll 1\, .
\end{equation}
Even if $\alpha _q$ is quite large, if the new factor $q_{\rm
now}/\Phi _{_{\rm N}}$ is small then the model can be compatible
with gravity. This is the thin shell effect.  It strongly depends
on the shape of the potential and, therefore, on the K\"ahler and
superpotential in the dark energy sector.

\section{Quintessential Puzzles}

Radiative corrections can modify the form of the quintessence
potential. In the Jordan frame where standard model matter couples
to $\tilde g_{\mu\nu}= A(q) g_{\mu\nu}$, the  quintessence field
only appears in the gravity part of the Lagrangian, i.e. the
Newton constant becomes $q$-dependent. Now, integrating out all
the standard model fields to obtain the effective action leads to
the appearance of a cosmological constant term $\Lambda_0^4$. No
contribution involving $q$ can appear as gravitational loops are
not taken into account. Going back to the Einstein frame implies
that the dark energy potential is modified by $\delta V_{\rm DE}=
\Lambda_0^4 A^4(q)$. The same result can be obtained using
the covariance of the action in the Einstein frame. Of course,
such a correction is huge as $A(q)= 1 + \dots$ \cite{P}. This is the usual
cosmological constant problem. Consistency imposes that
$\Lambda_0$ must be very small. In the following, we implicitly
assume that an unknown mechanism guarantees that $\Lambda_0=0$.

Let us come back to the structure of the scalar potential when the
quintessence superpotential is small compared to the hidden sector
superpotential $M\ll M_s$
\begin{equation}
V= V_{\rm DE}(Q) + \sum_i \vert F_{\rm z_i}\vert^2 + e^{\kappa
K}(K^{Q\bar Q} K_Q K_{\bar Q} -\frac{3}{\kappa}) \kappa^2 \vert
W\vert^2
\end{equation}
The first term $V_{\rm DE}$ contains terms of order $M^4$ and
$M_s^2 M^2$, it is responsible for the quintessence property of
the model. The second term contains the F-terms of the hidden
sector. The third term lead to a potential for the quintessence
field (if it does not vanish).

Let us consider first models where the Kahler potential can be
expanded around $Q=0$
\begin{equation}
K= Q\bar Q + \dots
\end{equation}
where $\dots$ represent Planck suppressed operators. The
quintessence field picks up a soft breaking mass\cite{CL,BM3}
\begin{equation}
V=V_{\rm DE} + m_{3/2}^2 \vert Q\vert^2
\end{equation}
where we must impose $\sum_i \vert F_{\rm z_i}\vert^2=
3m_{3/2}^2\kappa^{-1}$ in order to cancel the intolerably large
contribution to the cosmological constant coming from the hidden
sector. Due to the large value of $m_{3/2}$ compared to the
quintessence field, the potential acquires a minimum $Q_0$ small
in Planck units. The scale $M$ is tuned to get a minimum value for
the potential of order $\Omega_\Lambda \rho_c$. At this minimum,
the mass of the quintessence field is $m_{3/2}$, large enough to
evade all the gravitational tests. Now cosmologically, the
steepness of the quadratic potential in $Q$ implies that the field
must have settled at the minimum before BBN. If not the energy
density of the quintessence field would exceed the $\rm MeV$
energy scale of BBN. In practice, the potential is constant since
BBN, i.e. equivalent to a cosmological constant: a very intricate
manner of modelling a pure cosmological constant throughout most
of the universe history!

 One can circumvent this argument by taking singular
potentials where the potential term in $\vert W\vert^2$ is
constant. One can choose
\begin{equation}
K=-\frac{n}{\kappa} \ln \kappa^{1/2}(Q+\bar Q)
\end{equation}
In this case, n=3 for moduli and n=1 for the dilaton. Fine-tuning
of the cosmological constant requires
\begin{equation}
\sum_i \vert F_{\rm z_i}\vert^2= (3-n) m_{3/2}^2\kappa^{-1}
\end{equation}
leaving
\begin{equation}
V=V_{\rm DE}
\end{equation}
No mass term appears for the quintessence field. The mass of the
quintessence field at the minimum of the matter-dependent
potential is of order $H_0$. Moreover the thin-shell effect is
only present for small values of the normalised scalar field $q$.
This is not the case for well-motivated superpotentials motivated
such as the ones obtained from gaugino condensation. However, this is not
excluded for clever choices of the dark energy superpotential.

In conclusion, coupling dark energy to supersymmetry breaking
modifies  runaway potentials in a drastic way, giving a large mass
to the quintessence field of order of the gravitino mass. This can
only be avoided using no scale models. In this case, only very
special superpotentials can lead to a chameleon effect, and
therefore viable models. The construction of such models is
challenging and worth pursuing.

\bigskip
I would like to thank my long time friend and collaborator Jerome
Martin.

\def\Discussion{
\setlength{\parskip}{0.3cm}\setlength{\parindent}{0.0cm}
     \bigskip\bigskip      {\Large {\bf Discussion}} \bigskip}
\def\speaker#1{{\bf #1:}\ }
\def\endDiscussion{}


\begin{thebibliography}{99}

\bibitem{IA} S.~Perlmutter S {\it et al.}, Astrophys.~J. {\bf 517},
565 (1999); P.~M.~Garnavich {\it et al.}, Astrophys.~J. {\bf 493},
L53 (1998); A.~G.~Riess {\it et al.}, Astron.~J. {\bf 116}, 1009
(1998); P.~Astier {\it et al}, Astron.~Astrophys. {\bf 447}, 31
(2006).


\bibitem{RP} B.~Ratra and P.~J.~E.~Peebles, Phys.~Rev.~D {\bf 37},
3406 (1998).

\bibitem{quint} P.~G.~Ferreira and M.~Joyce, Phys.~Rev.~D {\bf 58},
023503 (1998).

\bibitem{ed} E.~Copeland,~M.~Sami and S.~Tsujikawa,
Int.~J.~Mod.~Phys.~D15 (2006) 1753

\bibitem{PB} P.~Bin\'etruy, Phys.~Rev.~D {\bf 60}, 063502 (1998), {\tt
hep-ph/9810553}; P.~Bin\'etruy, Int.~J.~Theor.~Phys. {\bf 39},
1859 (2000).

\bibitem{BMcosmo} P.~Brax and J.~Martin, Phys.~Lett. {\bf B468}, 40
(1999).


\bibitem{Nilles} H.~P.~Nilles, Phys.~Rept. {\bf 101}, 1 (1984);
S.~P.~Martin, {\tt hep-ph/9709356}; I.~J.~R.~Aitchison, {\it
Supersymmetry and the MSSM: An Elementary Introduction}, Notes of
Lectures for Graduate Students in Particle Physics, Oxford, 1004
(2005).







\bibitem{BM1}
Ph.~Brax and J.~Martin, Phys.~Lett.~{\bf B647} 320 (2007).

\bibitem{BM2}
Ph.~Brax and J.~Martin, JCAP 0611:008 (2006).
\bibitem{BM3}
Ph.~Brax and J.~Martin, Phys.~Rev.~ {\bf D 75} 083507 (2007).

\bibitem{GR} C.~M.~Will, Living.~Rev.~Rel. {\bf 9}, 2 (2006); E.~Fischbach and C.~Talmadge, {\it The Search for non-Newtonian
Gravity}, {\it Springer-Verlag, New-York}, (1999); B.~Bertotti,
L.~Iess and P.~Tortora, Nature {\bf 425}, 374 (2003);
G.~Esposito-Farese.

\bibitem{DP} T.~Damour and A.~M.~Polyakov, Nucl.~Phys.~{\bf B423}, 532
  (1994).

\bibitem{KW} J.~Khoury and A.~Weltman, Phys.~Rev.~D~{\bf 69}
(2004) 044026
\bibitem{cham} P.~Brax, C.~van~de~Bruck, A.~C.~Davis, J.~Khoury and
A.~Weltman, Phys.~Rev.~D {\bf 70}, 123518 (2004).
\bibitem{P} M.~Pietroni, Phys.~Rev.~D~{\bf 72} (2005) 043535.
\bibitem{CL} S.~Carroll and D.~Lyth, Phts.~Lett.~{bf B458} (1999)
197.



\end{thebibliography}
\end{document}